\documentclass[final,5p,times,twocolumn]{elsarticle}

\usepackage{amssymb}
\usepackage{amsmath}
\usepackage{bm}
\usepackage{subfig}
\usepackage{xcolor}
\usepackage{makecell}
\usepackage{float}
\journal{Journal of Computational Science}

\begin{document}

\begin{frontmatter}

%% Title, authors and addresses

%% use the tnoteref command within \title for footnotes;
%% use the tnotetext command for theassociated footnote;
%% use the fnref command within \author or \address for footnotes;
%% use the fntext command for theassociated footnote;
%% use the corref command within \author for corresponding author footnotes;
%% use the cortext command for theassociated footnote;
%% use the ead command for the email address,
%% and the form \ead[url] for the home page:
%% \title{Title\tnoteref{label1}}
%% \tnotetext[label1]{}
%% \author{Name\corref{cor1}\fnref{label2}}
%% \ead{email address}
%% \ead[url]{home page}
%% \fntext[label2]{}
%% \cortext[cor1]{}
%% \address{Address\fnref{label3}}
%% \fntext[label3]{}

\title{Cubature rules for weakly and fully compressible off-lattice Boltzmann methods}

%% use optional labels to link authors explicitly to addresses:
\author[us,hbrs]{Dominik Wilde} \corref{cor1}
\author[fub]{Andreas Krämer}
\author[hbrs,us]{Mario Bedrunka}
\author[hbrs,scai]{Dirk Reith}
\author[us]{Holger Foysi}

\cortext[cor1]{Corresponding author: wilde.aerospace@gmail.com}

\address[us]{Chair of Fluid Mechanics, University of Siegen, Paul-Bonatz-Straße 9-11, 57076 Siegen-Weidenau, Germany}
\address[hbrs]{Institute of Technology, Resource and Energy-efficient Engineering (TREE),\\ Bonn-Rhein-Sieg University of Applied Sciences,
Grantham-Allee 20, 53757 Sankt Augustin, Germany}
\address[fub]{Department of Mathematics and Computer Science, Freie Universität Berlin, Arnimallee 6, 14195 Berlin, Germany}
\address[scai]{Fraunhofer Institute for Algorithms and Scientific Computing (SCAI), Schloss Birlinghoven, 53754 Sankt Augustin, Germany}

\begin{abstract}
Off-lattice Boltzmann methods increase the flexibility and applicability of lattice Boltzmann methods by decoupling the discretizations of time, space, and particle velocities. However, the velocity sets that are mostly used in off-lattice Boltzmann simulations were originally tailored to on-lattice Boltzmann methods. In this contribution, we show how the accuracy and efficiency of weakly and fully compressible semi-Lagrangian off-lattice Boltzmann simulations is increased by velocity sets derived from cubature rules, i.e. multivariate quadratures, which have not been produced by the Gauß-product rule. In particular, simulations of 2D shock-vortex interactions indicate that the cubature-derived degree-nine D2Q19 velocity set is capable to replace the Gauß-product rule-derived D2Q25. Likewise, the degree-five velocity sets D3Q13 and D3Q21, as well as a degree-seven D3V27 velocity set were successfully tested for 3D Taylor-Green vortex flows to challenge and surpass the quality of the customary D3Q27 velocity set. In compressible 3D Taylor-Green vortex flows with Mach numbers $\mathrm{Ma}={\{0.5;1.0;1.5;2.0 \}}$ on-lattice simulations with velocity sets D3Q103 and D3V107 showed only limited stability, while the off-lattice degree-nine D3Q45 velocity set accurately reproduced the kinetic energy provided by literature.
\end{abstract}

\begin{keyword}
%% keywords here, in the form: keyword \sep keyword
%% PACS codes here, in the form: \PACS code \sep code

%% MSC codes here, in the form: \MSC code \sep code
%% or \MSC[2008] code \sep code (2000 is the default)
Lattice Boltzmann method \sep Cubature \sep Semi-Lagrangian \sep Gauss-Hermite quadrature \sep Compressible
\end{keyword}

\end{frontmatter}

%% \linenumbers

\section{Introduction}

The lattice Boltzmann method (LBM) \cite{McNamara1988,Higuera1989,Chen1991,Kruger2016,Lallemand2020} is an efficient approach for the numerical simulation of fluids. Compared to other methods discretizing the Boltzmann equation, such as discrete-velocity models \cite{Platkowski1988,Luo2000} or (unified) gas-kinetic schemes \cite{Su1999,Xu2010}, the LBM exhibits three key properties. First, the distribution functions are integrated along characteristics in time, leading to the well-known 0.5-shift in the relaxation time \cite{He1998a}. Second, the equilibrium distribution function is expressed as Hermite series, mostly of degree two, but higher for compressible \cite{Coreixas2017} and thermal cases \cite{Shan1998}. Third, the velocity space is integrated via quadrature rules, in particular by Gauß-Hermite quadratures \cite{Shan1998,Shan2006}. By the latter, the unbounded velocity space of the Boltzmann equation is expressed by only a limited number of weighted particle velocities, tied together as a velocity set. The type of numerical integration with the underlying weight function $\mathrm{exp}\left(-|\mathbf{x}|^2\right)$ does not only appear in the LBM, but for instance also in Kalman filters \cite{Arasaratnam2009}. Therefore, the literature on numerical integration is a good starting point for the search of well-suited velocity sets \cite{Stroud1973,Cools1993, Cools2003}.  More specifically,  the literature specifies two main approaches to derive the respective rules: Gauß-product rules and non-product rules \cite{Stroud1973}. The literature denotes multivariate non-product rules also as \emph{cubature rules} \cite{Cools1993,Cools2003}.

The simpler and widely-used approach of the above-mentioned ones is the Gauß-product rule, which is calculated by the outer product of a 1D quadrature. For example, the D1Q3 velocity set is essentially the 1D Gauß-Hermite quadrature constructed by the roots of the third order Hermite polynomial \cite{He1997}. By applying the Gauß-product rule, the D1Q3 is turned into a D2Q9 in two dimensions, and by applying the same rule once more the D3Q27 velocity set is derived \cite{He1997}. Since the number of particle velocities exceeds the number of the encoded physical moments, pruning the D3Q27 velocity set leads to the D3Q15 and D3Q19 sets with different high-order truncation errors \cite{White2011,Silva2014,Bauer2020}. A modification of the regular D3Q27 is the recently introduced crystallographic LBM \cite{Namburi2016}, which expresses the particle distributions in a body centered cubic arrangement of grid points.  All above mentioned velocity sets fit into a regular grid, which is another key asset of the well-established on-lattice Boltzmann method, with the vector of the particle velocities ending on one of the neighbouring grid points. For weakly compressible flows, this collection of velocity sets in combination with a second-order polynomial expansion of the Maxwell-Boltzmann equilibrium is widely used, although not perfect in terms of Galilean invariance, due to well-known errors in the stress tensor \cite{Qian1993}. However, when a higher degree of precision for the numerical integration is needed \cite{Philippi2006, Shan2016}, in particular for compressible \cite{Frapolli2015} or dilute flows \cite{Meng2011}, the resulting velocity sets become unfeasible, for two reasons. Firstly, Gauß-product rules suffer from the "curse of dimensionality", i.e. the number of abscissae rapidly increases especially for high spatial dimensions with high degree of precision. Secondly, when derived from Hermite polynomials the degree-nine one-dimensional velocity set D1Q5 holds abscissae, which do not fit on a regular grid, so that an even larger velocity set D1Q7 with equidistant abscissae must be used to obtain a sufficiently high integration order \cite{Chikatamarla2009}. These multivariate lattices with equidistant nodes are constructed by solving the orthogonality relations  \cite{Shan2006,Philippi2006,Shan2010}. 

Contrary to Gauß-product rules, the abscissae of cubature rules are not derived by quadratures of lower dimension. They are rather found individually for a certain pair of degree of precision and dimension. Due to this freedom of velocity discretization, off-lattice Boltzmann methods are required to apply them in the LBM. There is only a very limited number of publications, which display simulations using off-lattice velocity sets \cite{Surmas2009,Yudistiawan2010a,Tamura2011}. 

The present manuscript therefore derives velocity sets from cubature rules and explores them using the semi-Lagrangian lattice Boltzmann method (SLLBM, \cite{Kramer2017}),  with applications to both weakly and fully compressible flows.  In contrast to Eulerian time integration schemes like finite difference \cite{Cao1997, Hejranfar2017}, finite volume \cite{Nannelli1992,Li2016}, discontinuous Galerkin LBM schemes\cite{Shi2003,Min2011}, the SLLBM inherits the Lagrangian time integration along characteristics from the LBM and recovers the off-lattice distribution function values by interpolation. Recent works \cite{DiIlio2018, Dorschner2018, Saadat2019, Reyhanian2020, Wilde2020, Kramer2020, Saadat2020} provided evidence that the SLLBM is a promising off-lattice Boltzmann method for the simulation of both weakly and compressible flows. However, these works used either on-lattice velocity sets, e.g. D2Q9, or D3Q27, or velocity sets that had been derived by the Gauß-product rule, e.g. D2Q25 from 1D Gauß-Hermite quadrature. This gap is closed by the present article. This work shows that cubature-based velocity sets significantly enhance off-lattice Boltzmann simulations in terms of both efficiency and accuracy. 

The remainder of this manuscript is structured as follows. Quadrature and cubature in the LBM are briefly recapitulated in section \ref{sec:cubature} with a list of all investigated velocity sets in this article. Section \ref{sec:model} details the semi-Lagrangian lattice Boltzmann model for both weakly and fully compressible flows. The results section \ref{sec:results} studies three test cases: a compressible two-dimensional shock-vortex interaction and the three-dimensional Taylor-Green vortex both in the weakly and in the fully compressible regime, each of them requiring different equilibria and velocity sets. Sections \ref{sec:discussion} and \ref{sec:conclusion} provide discussion and conclusion.

\section{Cubature in lattice Boltzmann methods} \label{sec:cubature}
Following \cite{Cools2003}, we consider the approximation of an integral of function $\mathcal{F}$
\begin{equation}\label{eq:integral}
    I(\mathcal{F}) = \int_\Omega \omega(\mathbf{x})\mathcal{F} (\mathbf{x}) d\mathbf{x},
\end{equation}

with $\Omega \subset \mathbb{R}^D$, weight function $\omega(\mathbf{x}) \geq 0$ and dimensions $D\geq2$ by quadrature or cubature rules of the form

\begin{equation}
    \mathcal{C}(\mathcal{F}) = \sum_{i=0}^{Q-1} w_i \mathcal{F}(\mathbf{x}_i), 
\end{equation}

with number of abscissae $Q$ and discrete weights $w_i$.

The degree of a quadrature is defined as the largest integer $d$ that yields $I(\mathcal{F}) = \mathcal{C}(\mathcal{F}) $ for all monomials
\begin{equation}
    \prod_{i=0}^{D-1} x_i^{j_i} \quad \mathrm{with} \quad \sum_{i=0}^{D-1} j_i \leq d.
\end{equation}
of degree $\leq d$.

These cubature formulas are applied to the distribution function of the force-free BGK-Boltzmann equation 
\begin{equation}\label{eq:Boltzmann}
    \frac{\partial f }{\partial t} + \boldsymbol \xi \cdot \nabla f= -\frac{1}{\lambda}\left(f -f^\mathrm{eq}\right),
\end{equation} 

with (unshifted) relaxation time $\lambda$,  particle distribution function $f$,  equilibrium distribution function $f^\mathrm{eq}$, and particle velocities $\boldsymbol{\xi}$. To that end, the Hermite moments $\boldsymbol{a}^{(n)}$ are gained in the form \cite{Shan1998,Shan2006}

\begin{equation}\label{eq:Hermite_moments}
     \boldsymbol{a}^{(n)} = \int_{\mathbb{R}^D} f \boldsymbol{\mathcal{H}}^{(n)} d\boldsymbol{\xi},
\end{equation}
where $\boldsymbol{\mathcal{H}}^{(n)}$ is a nth-order Hermite polynomial. The distribution function $f$ is expressed as a finite Hermite series

\begin{equation}
    f \approx f^N = \omega(\boldsymbol{\xi}) \sum_{n=0}^N \frac{1}{n!}\boldsymbol{a}^{(n)}  \boldsymbol{\mathcal{H}} ^{(n)},
\end{equation}
with  truncation order $N$ \cite{Shan2006} with the result that moments of order $M$ are exactly represented, if $M \leq N$. 
The velocity space is discretized by replacing the integral of Eq. \eqref{eq:Hermite_moments} by a weighted quadrature leading to
\begin{align} \label{eq:quadrature}
    \boldsymbol{a}^{(n)} &= \int_{\mathbb{R}^D} \omega(\boldsymbol{\xi}) \frac{f}{\omega(\boldsymbol{\xi})} \boldsymbol{\mathcal{H}}^{(n)}d\boldsymbol{\xi} \\ &= \sum_{i=0}^{Q-1} \frac{w_i f(\boldsymbol{\xi}) \boldsymbol{\mathcal{H}}^{(n)}(\boldsymbol{\xi}_i)}{\omega(\boldsymbol{\xi}_i)} = \sum_{i=0}^{Q-1} {f_i \boldsymbol{\mathcal{H}}^{(n)}_i}, 
\end{align}
with the replacements $f_i = w_i f(\boldsymbol{\xi}) /\omega(\boldsymbol{\xi}_i)$ and $\boldsymbol{\mathcal{H}}^{(n)}(\boldsymbol{\xi}_i)=\boldsymbol{\mathcal{H}}^{(n)}_i$. From \cite{Nie2008} it is known that moments of order $M$ can be exactly determined by quadrature or cubature rules of degree $d \geq N~+~M$. In particular, for weakly compressible flows the order of the moments is usually limited to $N=2$, although $N=3$ cancels the cubic error when deriving the Navier-Stokes equations by a Chapman-Enskog analysis \cite{Shan2006}. In contrast, fully compressible flows require even expansion order $N=4$. By overcoming the restriction that abscissae $\boldsymbol\xi_i$ need to match a regular grid, cubature rules become applicable, provided they approximate the weight function
\begin{equation}
    \omega(\boldsymbol\xi) = \frac{1}{(2\pi)^{D/2}}e^{-|\boldsymbol{\xi}|^2/2}.
\end{equation}
In the literature, cubature rules are usually listed for the weight function $\Tilde{\omega}(x) = \mathrm{exp}\left(-|\mathbf{x}|^2\right)$. To obtain a velocity set to approximate the moments in Eq. \eqref{eq:quadrature}, the cubature's abscissae have to be scaled by $\sqrt{2}$. The lattice speed of sound $c_s$ of these velocity sets is---unless otherwise specified---unity $c_s=1$, which also implies the reference temperature $T_0=1$. 

\begin{figure}
    \centering
    \includegraphics{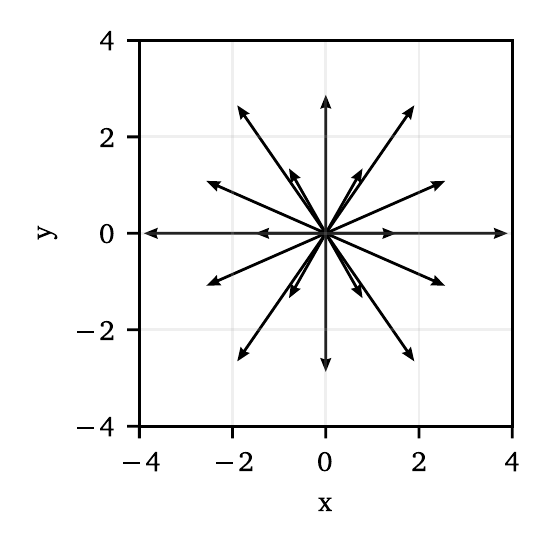}
    \caption{Shape of the D2Q19 velocity set.}
    \label{fig:d2q19}
\end{figure}

\begin{figure}
\centering
\subfloat{  
    \includegraphics[width=0.3\linewidth]{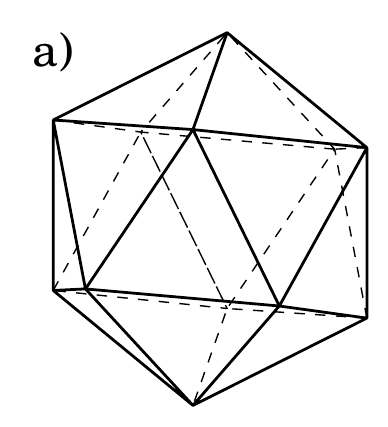}
}
\subfloat{    
\includegraphics[width=0.33\linewidth]{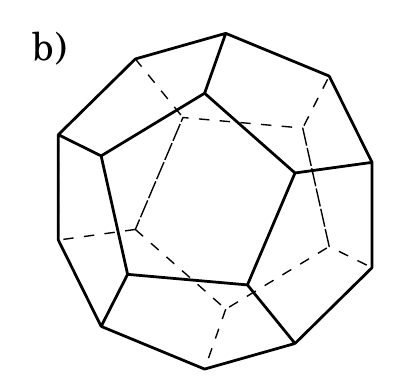}
}
    \caption{Shape of icosahedron a) and dodecahedron b), which are the bases for the velocity sets D3Q13 and D3Q21 used in this work.}
    \label{fig:platonics}
\end{figure}

The following velocity sets were identified to be subject of an in-depth examination: 
\begin{itemize}
    \item First, the newly introduced degree-nine D2Q19 in comparison to the D2Q25 derived from Gauß-product rule for fully two-dimensional compressible flows. The shape of D2Q19 is shown in Fig. \ref{fig:d2q19}
    \item Second, in three dimensions velocity sets for both weakly and fully compressible flows were chosen. The D3Q13 based on a icosahedron was already used in a weakly compressible finite difference LBM with a triangular mesh \cite{Tamura2011}, whereas the D3Q21 derived from a dodecahedron has not been examined so far. Both platonic solids, shown in Fig.~\ref{fig:platonics}, possess a high geometric isotropy, i.e. the flow information encoded into the distribution function values is transported by regularly spaced abscissae ending on the surface of a sphere.  The weights and abscissae of the D3Q13 and D3Q21 velocity sets are listed in Tables \ref{tab:d3q13} and \ref{tab:d3q21}, respectively. Both velocity sets were compared to the customary D3Q27 on-lattice velocity set and in addition to a degree-seven velocity set, which we call D3V27 \cite{Stroud1973}. This velocity set is not affected by the cubic error in the stress tensor that troubles degree-five velocity sets. Although in the same spirit, it is not identical to the velocity set presented by Yudistiawan et al. \cite{Yudistiawan2010a}.
    \item   Last, for three-dimensional compressible flows, a recently introduced D3Q45 velocity set was applied \cite{Wilde2020a}, based on a cubature rule by Konyaev \cite{konyaev1977,VanZandt2019}. This off-lattice velocity set was compared in an off-lattice Boltzmann simulation to the state of the art on-lattice counterparts D3Q103 and D3V107 applied to compressible on-lattice Boltzmann simulations. 
\end{itemize}

Table \ref{tab:sets} designates the velocity sets, which were used for the simulations in this article. The additional notation $E_{D,d}^Q$ follows the literature on multivariate quadratures integrating the weight function $\Tilde{\omega}(x) = \mathrm{exp}\left(-|\mathbf{x}|^2\right)$, listing dimension $D$, degree $d$ and number of abscissae $Q$. To identify the suitability of the introduced velocity sets beforehand, the symmetry conditions in \ref{app:symmetry} were successfully tested \cite{Frisch1986}. For example, a degree-five velocity set only reproduces the symmetry condition up to fifth order, whereas degree-nine velocity sets will also correctly reproduce the symmetry conditions up to ninth order.

\begin{table}
    \centering
    \caption{Quadrature and cubature rules utilized in the present article. Notation $E_{D,n}^Q$ according to Stroud \cite{Stroud1973} with number of abscissae~$Q$, dimension~$D$, and degree of precision~$d$.}
    \label{tab:sets}
\begin{tabular}{ c p{0.08\linewidth} p{0.4\linewidth} c } 
  Name & $E_{D,d}^Q$ & Remarks & Sources  \\
  \hline
  D2Q19 & $E_{2,9}^{19}$ & Cubature rule by Haegemans and Piessens & \makecell{\cite{Haegemans1977},\\ \ref{app:d2q19}}\\
  D2Q25 & $E_{2,9}^{25}$ & Gauß-product rule from 1D degree-nine Gauß-Hermite quadrature & \cite{Chikatamarla2009} \\ 
  D3Q13 & $E_{3,5}^{13}$ & Derived by the vertices of an icosahedron & \makecell{\cite{Stroud1973,Tamura2011},\\ Table \ref{tab:d3q13}}\\
  D3Q21 & $E_{3,5}^{21}$ & Derived by the vertices of a dodecahedron & \makecell{\cite{Stroud1973},\\ Table \ref{tab:d3q21}} \\
  D3Q27 & $E_{3,5}^{27}$ & Doubly applied Gauß-product rule from 1D Gauß-Hermite quadrature & \cite{Stroud1967}\\
  D3V27 & $E_{3,7}^{27}$ & Cubature rule by Stroud and Secrest. & \cite{Stroud1973} \\  
  D3Q45 & $E_{3,9}^{45}$ & Cubature rule by Konyaev. & \makecell{\cite{konyaev1977,VanZandt2019}, \\ \ref{app:d3q45}}\\
  D3Q103 & $E_{3,9}^{103}$ & On-lattice velocity set & \makecell{\cite{Shan2016}} \\ 
  D3V107 & $E_{3,9}^{107}$ & On-lattice velocity set & \cite{Surmas2009}

\end{tabular}
\end{table}

\begin{table}
    \centering
    \caption{Abscissae $\xi_i$ and weights $w_i$ for the D3Q13 velocity set, based on an icosahedron \cite{Stroud1973,Tamura2011,Kruger2016} with  $r^2=(5+\sqrt{5})/2$ and $s^2=(5-\sqrt{5})/2$. The lattice speed of sound is $c_s=1$.}
    \label{tab:d3q13}
\renewcommand{\arraystretch}{1.5}
\setlength{\tabcolsep}{7pt}
\begin{tabular}{ c c c } 
  $i$ & $w_i$ &  $\xi_i$ \\
  \hline
  $0$ & $2/5$ & $(0,0,0)$  \\
  $1,\ldots,4$ &$1/20$& $(0,\pm r,\pm s)$  \\ 
  $5,\ldots,8$ & $1/20$ & $(\pm s, 0,\pm r)$  \\
  $9,\ldots,12$ & $1/20$  & $(\pm r,\pm s, 0)$ \\ 
\end{tabular}
\end{table}

\begin{table}
    \centering
    \caption{Abscissae $\xi_i$ and weights $w_i$ for the D3Q21 velocity set, based on a dodecahedron \cite{Stroud1973} with $\phi = (1 + \sqrt{5}) / 2$. The lattice speed of sound $c_s$ is $c_s=\sqrt{3/5}$.}
    \label{tab:d3q21}
\renewcommand{\arraystretch}{1.5}
\setlength{\tabcolsep}{7pt}
\begin{tabular}{ c c c } 
  $i$ & $w_i$ & $\xi_i$ \\
  \hline
  $0$ & $2/5$ & $(0,0,0)$  \\
  $1,\ldots,8$ & $3/100$ & $(\pm 1,\pm 1,\pm 1)$  \\
  $9,\ldots,12$ & $3/100$ & $(0,\pm \phi,\pm 1/\phi)$  \\ 
  $13,\ldots,16$ & $3/100$ & $(\pm 1/\phi, 0,\pm \phi)$ \\
  $17,\ldots,20$ & $3/100$ & $(\pm \phi,\pm 1/\phi, 0)$  \\ 
\end{tabular}
\end{table}

\section{Model description} \label{sec:model}

The lattice Boltzmann equation reads
\begin{equation}
    f_i(\mathbf{x}+\boldsymbol{\xi}_i\delta_t,t+\delta_t) = f_i(\mathbf{x},t) - \frac{1}{\tau}\left(f_i(\mathbf{x},t)-f_i^\mathrm{eq}(\mathbf{x},t)\right)
\end{equation}
with relaxation time $\tau = \nu /(c_s^2 \delta_t) + 0.5$ depending on kinematic viscosity $\nu$, and time step size $\delta_t$.

This work applies the discrete equilibrium distribution function based on an expansion in terms of Hermite polynomials

\begin{equation}    f_i^{\mathrm{eq},N}(\mathbf{x},t) = w_i \sum_{n=0}^N \frac{1}{n!}\boldsymbol{a}^{(n)}_{eq}(\mathbf{x},t) :  \boldsymbol{\mathcal{H}}_i ^{(n)},
\end{equation}
with $"\!\!:\!\!"$ denoting full contraction, $w_i$ the discrete weights and $N$ the expansion order. Both the moments $\boldsymbol{a}^{(n)}_{eq}$ and the Hermite tensors $\boldsymbol{\mathcal{H}}_i^{(n)}$ are listed in the Appendix. In the weakly compressible case the local temperature $\theta = T / T_0$ is set to $\theta=1$ with vanishing terms for second and higher order moments.  
With the help of Eq. \ref{eq:quadrature} the moments density and momentum are obtained 
\begin{align}
\rho&=\sum_{i=0}^{Q-1} f_i \label{eq:density}\\
\rho \mathbf{u} &= \sum_{i=0}^{Q-1} \boldsymbol{\xi}_{i} f_i \label{eq:momentum}.
\end{align} 
Compared to the standard LBM, the SLLBM replaces the node-to-node streaming step by a cell-wise interpolation procedure using interpolation polynomials $\psi$ of interpolation order $p$ to obtain the departure points, whose locations are dictated by the respective reversed particle velocities $f_i(\mathbf{x}-\delta_t \boldsymbol{\xi}_i)$ , i.e. for all $\mathcal{M}$ points $\mathbf{x}$ in each cell $\Xi$:
\begin{equation}\label{eq:sl_streaming}
f_i(\mathbf{x},t) = \sum_{j=1}^{\mathcal{M}} \hat{f}_{i\Xi j}(t) \psi_{\Xi j}(\mathbf{x}),
\end{equation}

with $\hat{f}_{i\Xi j}$ denoting the distribution function value at the support points in each cell.

Gauß-Lobatto-Chebyshev points were used for the distribution of support points in the cells \cite{Wilde2020,Kramer2017}.

For the compressible test cases in Sections \ref{sec:SVI} and \ref{sec:TGV3Dfully} we employed the recently introduced compressible SLLBM \cite{Wilde2020} with the following extensions. A second distribution function $g$ is introduced to enable the calculation of flows with variable heat capacity ratio $\gamma$ \cite{Nie2008}, following the same collide-and-stream algorithm as applied to the distribution function $f$. 
\begin{equation}
    g_i(\mathbf{x}+\boldsymbol{\xi}_i\delta_t,t+\delta_t) = g_i(\mathbf{x},t) - \frac{1}{\tau}\left(g_i(\mathbf{x},t)-g_i^\mathrm{eq}(\mathbf{x},t)\right)
\end{equation}
The equilibrium of g is determined by the relation
\begin{equation}
    g^\mathrm{eq}_i = \theta (2 C_v - D) f^\mathrm{eq}_i,
\end{equation}
with heat capacity at constant volume $C_v$.
To complement Eqs. \eqref{eq:density} and \eqref{eq:momentum}, the local temperature $\theta$ is  determined by
\begin{equation}
     2\rho C_v \theta = \sum_{i=0}^{Q-1} \left(|\boldsymbol{\zeta}_i|^2 \frac{f_i}{T_0}  \!+g_i \right),
\end{equation}
with peculiar particle velocity $\boldsymbol\zeta_{i} = \boldsymbol\xi_{i} - \mathbf{u}$. \\
The local speed of sound $c_s^*$ consequently depends on the local relative temperature $\theta$ and on the heat capacity ratio $\gamma$, via 
\begin{equation}
    c_s^* = c_s \sqrt{\theta \gamma}.
\end{equation}
The relaxation time $\tau$  in the compressible case is dependent on the local pressure $P$, i.e. $\tau= \mu/(c_s^2 \delta_t P) +0.5$ with dynamic viscosity $\mu$. In addition, a quasi-equilibrium approach was used to vary the Prandtl number $\mathrm{Pr}$. More details about the compressible SLLBM solver are listed in \cite{Wilde2020}. 
The NATriuM solver was used for the off-lattice Boltzman simulations \cite{Kramer2020}, which is based on the finite element library deal.ii \cite{dealII92}. The on-lattice Boltzmann simulations were performed using the lettuce software \cite{andreas_kramer_2020_3757641,Bedrunka2021}, being based on the GPU-accelerated machine learning toolkit PyTorch \cite{torch}.

\section{Results} \label{sec:results}
The present work considers three test cases to evaluate the introduced velocity sets: the two-dimensional shock-vortex interaction, comparing the degree-nine velocity sets D2Q19 and D2Q25. This flow challenges the solver by acoustic emissions by the vortex downstream the shock. Next, the weakly compressible three-dimensional Taylor-Green vortex \cite{Brachet1983} was simulated by various 3D velocity sets, e.g. D3Q13, D3Q21, D3V27 and D3Q27 to explore the capability of the velocity sets to deal with underresolved flows. Finally, simulations of fully compressible 3D Taylor-Green vortex flows were explored with Mach numbers 0.5, 1.0, 1.5, and 2.0. This test case exhibits shocklets and viscous effects and features a standardized initialization. The on-lattice simulations were performed using the D3Q103 and D3V107 velocity sets, while the off-lattice simulations were run by the D3Q45 velocity set.
\subsection{Compressible 2D shock vortex interaction}\label{sec:SVI}
\begin{figure}[h]
    \centering
    \includegraphics[width=0.9\linewidth]{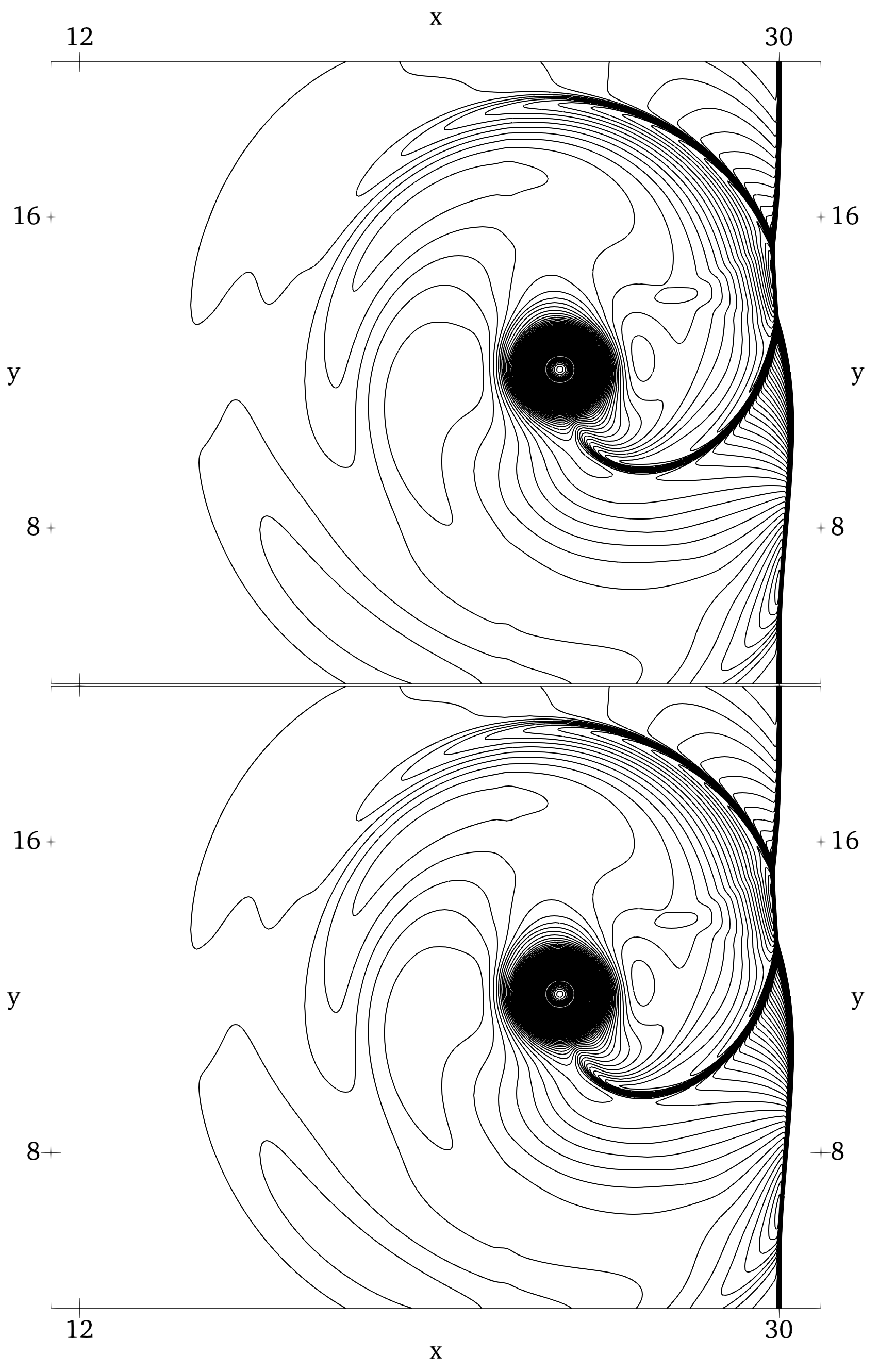}
    \caption{Density contours of the shock-vortex interaction with $Ma_v = 0.5$ and $Re=400$ in the range $\rho \in [0.92,1.55]$ in 119 steps with velocity sets D2Q19 (top) and D2Q25 (below). No significant difference is visible despite the D2Q19's 24 percent reduction in computational cost.}
    \label{fig:svi_contours}
\end{figure}

The compressible two-dimensional shock-vortex interaction was tested as a first test case to compare the two velocity sets D2Q19 and D2Q25. This benchmark was intensively studied by Inoue and Hattori \cite{Inoue1999a}, with the following setup. Two regions with $\mathrm{Ma}_a=1.2$ and $\mathrm{Ma}_b$ determined by Rankine-Hugoniot conditions are divided by a steady shock. The Reynolds number is defined as $\mathrm{Re}=c_{s,\infty}^* R / \nu_\infty$ with subscript $\infty$ denoting the inflow conditions. The flow field of the vortex is given by 
\begin{equation}
    u_\theta(r)=\sqrt{\gamma T}\mathrm{Ma_v}\,r\, \mathrm{exp}((1-r^2)/2),
\end{equation}
where $\mathrm{Ma}_v$ denotes the Mach number of the vortex. The initial pressure and density field are 
\begin{equation}
    P(r)=\frac{1}{\gamma}\left(1-\frac{\gamma-1}{2}\mathrm{Ma_v}\mathrm{exp}(1-r^2)\right)^{\gamma/(\gamma-1)}
\end{equation}
and 
\begin{equation}
    \rho(r)=\left(1-\frac{\gamma-1}{2}\mathrm{Ma_v}\mathrm{exp}(1-r^2)\right)^{1/(\gamma-1)}.
\end{equation}
The resolution was $256\times256$ cells with polynomial order $p=4$, ie. $1024\times1024$ grid points for the physical domain of $60R\times24R$. Initially the centre of the vortex was located at $x_\mathrm{v} = 32$, while the shock was located at $x_\mathrm{shock}= 30$. Likewise as in \cite{Wilde2020} we used a stretched grid to spatially resolve the shock region. The time step size was $\delta_t = 0.002$ with characteristic time $t'=R / c_{s,\infty}^*$. The Prandtl number was $\mathrm{Pr}=0.75$ in all cases. Table \ref{tab:setup-2Dvortex} designates the most important parameters of the simulations. Fig.~\ref{fig:svi_contours} depicts the density contours at $t=8$ for test case a) with $\mathrm{Ma_v}=0.5$, $\mathrm{Re}=400$ and provides evidence that there is no significant difference between the simulations run by the D2Q19 and D2Q25 velocity sets. 

\begin{table}
    \centering
    \caption{Simulation setup of the compressible 2D shock vortex interaction.}
    \label{tab:setup-2Dvortex}
\begin{tabular}{ l c c c} 
Test case & & a) & b) \\ 
\hline
  Reynolds number & $\mathrm{Re}$ & 400 & 800 \\
  Vortex Mach number & $\mathrm{Ma_{v}}$ & 0.5 & 0.25 \\
  Advection Mach number & $\mathrm{Ma_{a}}$ & 1.2 & 1.2 \\
  Relaxation time at $\theta=1.0$ & $\tau$  & 3.82 & 2.16 \\
  Minimal spatial resolution & $\delta_x$  & 0.003 & 0.003 \\ 
  Temporal resolution & $\delta_t$  & 0.002 & 0.002 \\
  Interpolation polynomial order & $p$ & 4 & 4\\
  Truncation order of equilibrium & $N$ & 4 & 4
\end{tabular}
\end{table}

\begin{figure}
    \centering
    \includegraphics{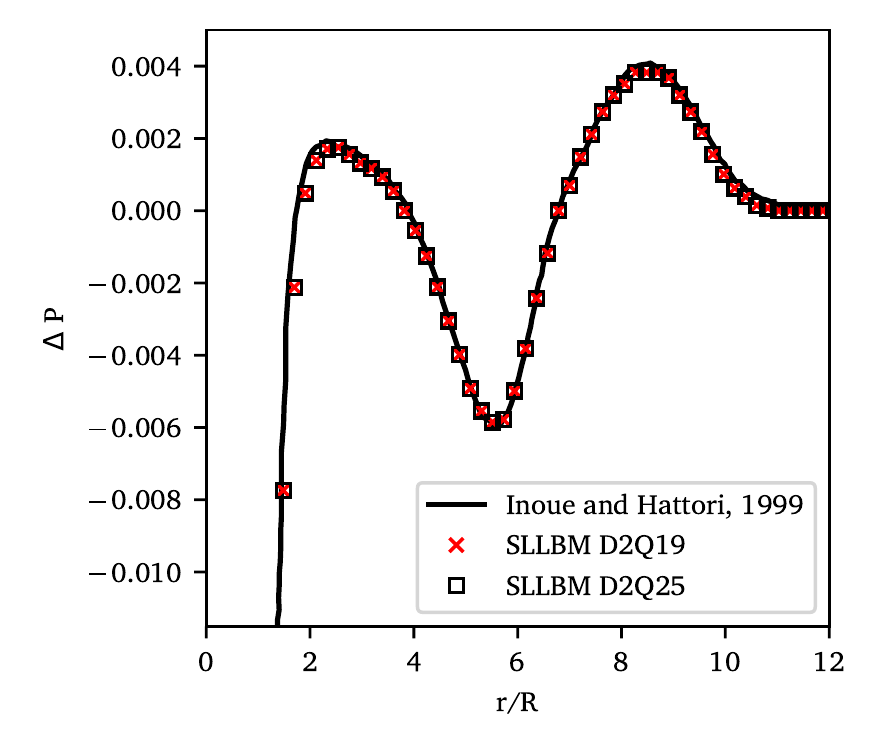}
    \caption{Comparison of D2Q19 and D2Q25 in terms of the sound pressure $\Delta P = (P - P_B) / P_B$ of the shock-vortex interaction with $Re=800, Ma_v=0.25$ and at time $t'=8'$ measured from the center of the vortex along the radius with $135^{\circ}$ with respect to the x-axis. The SLLBM match the reference by Inoue and Hattori \cite{Inoue1999a} well.}
    \label{fig:svi_pressure}
\end{figure}

 The evaluation of the sound pressure for test case b) at $\mathrm{Ma_v}=0.25$ and $\mathrm{Re}=800$ confirms this observation, being shown in Fig. \ref{fig:svi_pressure}. The sound pressure $\Delta P = (P - P_B) / P_B$, with subscript $B$ denoting the pressure downstream, was measured at time $t=8$ from the center of the vortex along the $135^\circ$ inclined radius with respect to the x-axis. As expected, the sound pressure of both SLLBM simulations coincided well with the reference solution by Inoue and Hattori \cite{Inoue1999a}. These findings approve the cubature-derived D2Q19 as a potent alternative for compressible off-lattice simulations to the D2Q25 being based on the Gauß-product rule used in previous works \cite{Dorschner2018,Wilde2020}. 
 
\subsection{Weakly compressible 3D incompressible Taylor-Green vortex} \label{sec:TGV3Dweakly}
The three-dimensional Taylor Green vortex served as a test case to apply the degree-five velocity sets D3Q13, D3Q21, and degree-seven D3V27 in comparison to the customary degree-five D3Q27 using an isothermal configuration $\theta=1$ with truncation order of the equilibrium $N=2$ for D3Q13, D3Q21, and D3Q27. The D3V27 simulation instead was equipped by a third truncation order  $N=3$. The Mach number was $Ma=0.1$. The computational domain is $S=[0,2\pi]^3$ and the following initial conditions were used
\begin{align}
\bm{u}(\bm{x},0) = \begin{pmatrix} \mathrm{sin}(x) \mathrm{cos}(y) \mathrm{cos}(z) \\  - \mathrm{cos}(x) \mathrm{sin}(y) \mathrm{cos}(z) \\ 0 \end{pmatrix},\label{eq:TGV3D-velocity}
\end{align}
\begin{equation}
    P(\bm{x},t=0) = \frac{1}{16} (\mathrm{cos}(2x) + \mathrm{cos}(2y)) \mathrm{cos}(2z+2),
\end{equation}
The kinetic energy of the flow is defined as
\begin{equation}
k = \frac{1}{(2 \pi)^3} \int_S \ \frac{1}{2} |\mathbf{u}|^2\ d^3 \mathbf{x}
\end{equation}
and the enstrophy as a measure of the dissipation with respect to the resolved scales \cite{Kramer2019}
\begin{equation}
\mathcal{E} = \frac{1}{(2 \pi)^3} \int_S \ (\nabla \times \mathbf{u})^2 \  d^3 \mathbf{x}.
\end{equation}
As explained in detail in \cite{Kramer2017} and references therein, the scaled enstrophy $\nu\mathcal{E}$ measures the physical dissipation in terms of the local shear stresses. In turn, its deviation from the phenomenological dissipation rate -dk/dt provides a measure for the numerical dissipation due to discretization. 

First, a well-resolved flow was tested by setting the Reynolds number to $Re=400$. The domain was discretized by $32^3$ cells with order of finite element $p=4$, resulting in $128^3$ grid points. The velocity sets were scaled in such a way that the time step size was $\delta t = 0.0015$ in all cases. A summary of the flow variables is listed in Table \ref{tab:setup-3Dvortex}.

\begin{table}
    \centering
    \caption{Simulation setup of the weakly compressible 3D Taylor-Green vortex.}
    \label{tab:setup-3Dvortex}
\begin{tabular}{ l c c c} 
Test case & & a) & b) \\ 
\hline
  Reynolds number & $\mathrm{Re}$ & 400 & 1600 \\
  Mach number & $\mathrm{Ma}$ & 0.1 & 0.1  \\
  Relaxation time & $\tau$  & 0.517 & 0.508 \\
  Spatial resolution & $\delta_x$  & 0.05 & 0.05 \\ 
  Temporal resolution & $\delta_t$  & 0.0015 & 0.00075\\
  Interpolation polynomial order & $p$ & 4 & 4\\
  Truncation order of equilibrium & $N$ & 2* & 2* \\
  \hline
  *$N=3$ for D3V27
\end{tabular}
\end{table}

\begin{figure}
    \centering
    \includegraphics{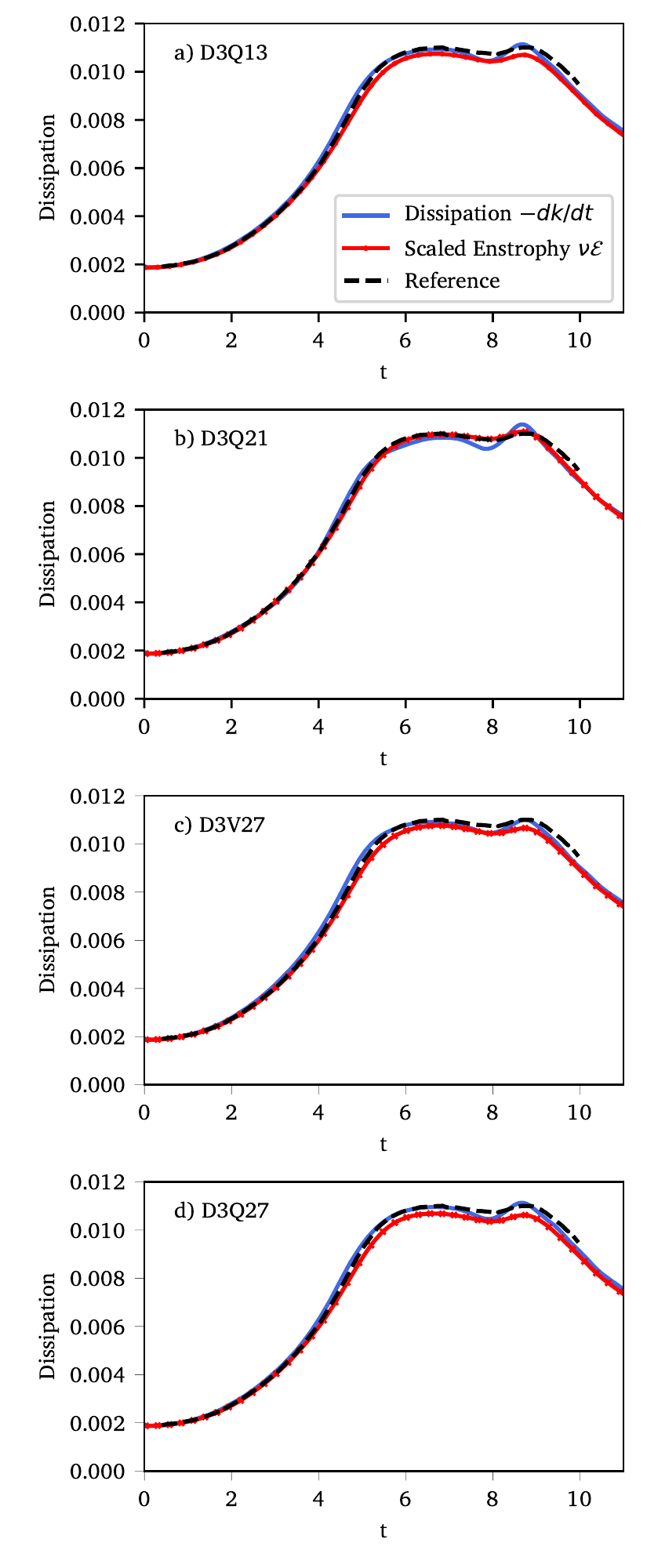}
    \caption{Dissipation and scaled enstrophy over time from simulations of the three-dimensional Taylor-Green vortex at Reynolds number $\mathrm{Re}=400$ using the velocity sets D3Q13 a), D3Q21 b), D3V27 c), and D3Q27 d). For this well-resolved simulation at $N=128³$ both the dissipation $-dk/dt$ and the scaled enstrophy $\nu \mathcal{E}$ matched the reference solution well for all velocity sets considered.}
    \label{fig:tgv_case1}
\end{figure}

\begin{figure}
    \centering
    \includegraphics{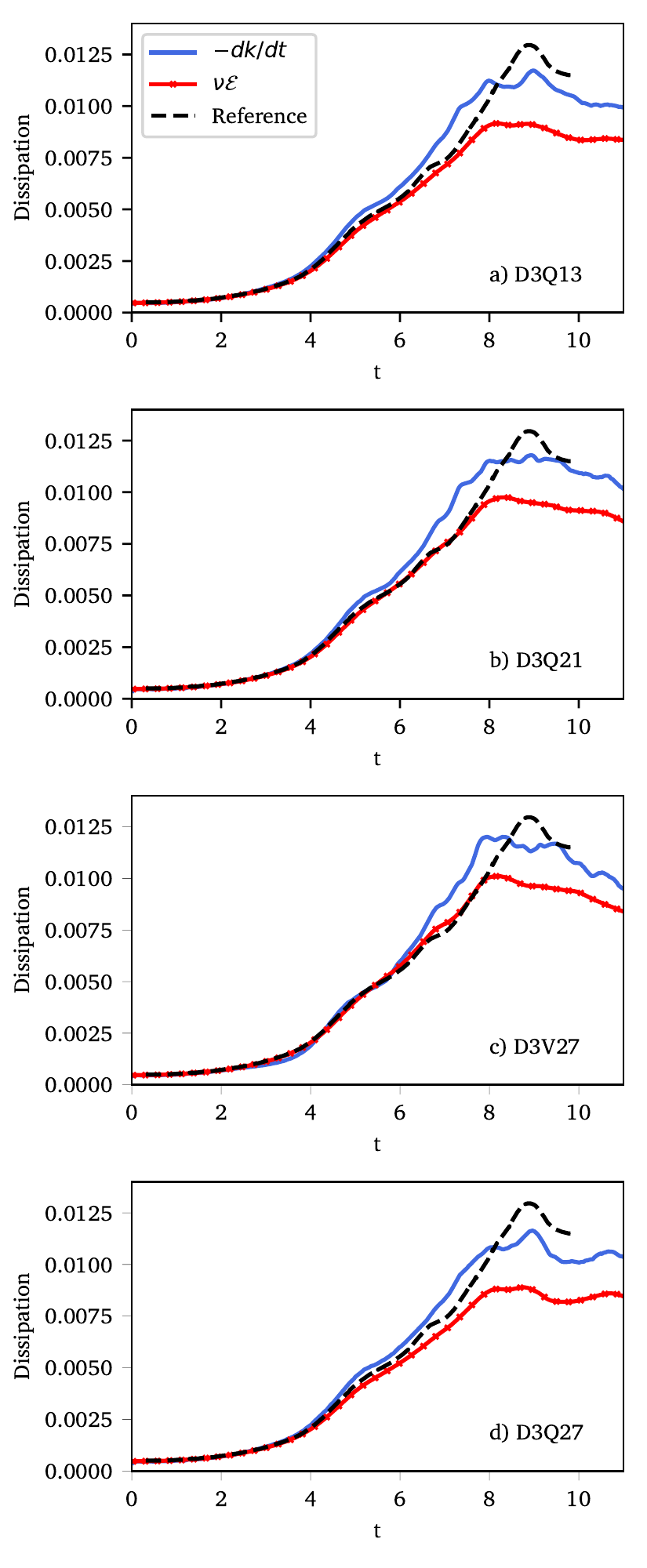}
    \caption{Dissipation $-kE/dt$ and scaled enstrophy over time from simulations of the three-dimensional Taylor-Green vortex at Reynolds number $\mathrm{Re}=1600$ using the velocity sets D3Q13 a), D3Q21 b), D3V27 c), and D3Q27 d). This under-resolved simulation at $N=128³$ reveals differences for the dissipation and scaled enstrophy for the respective velocity sets. }
    \label{fig:tgv_case2}
\end{figure}

\begin{figure}
    \centering
    \includegraphics{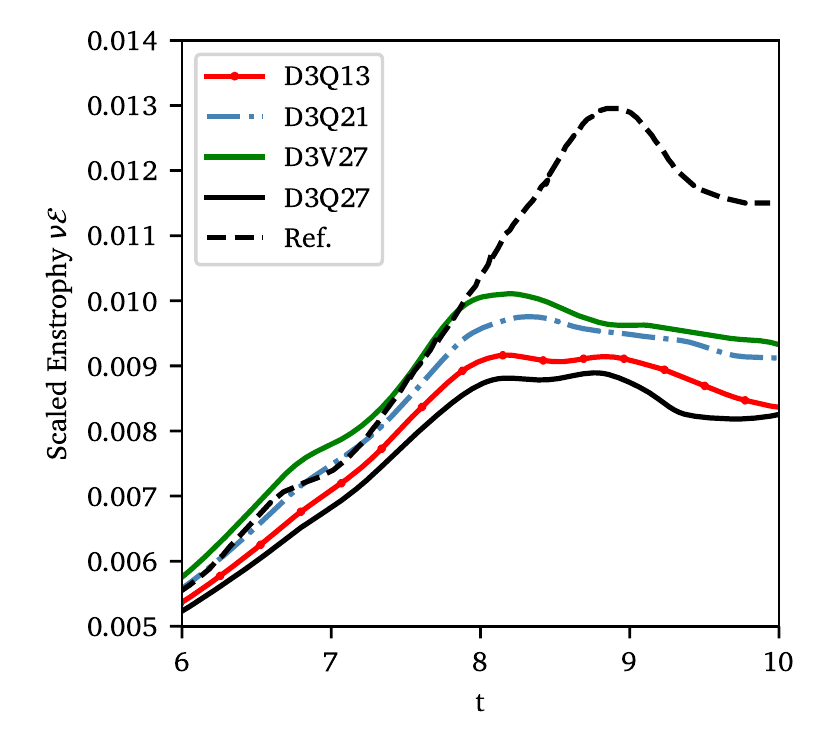}
    \caption{Comparison of the scaled enstrophies $\nu \mathcal{E}$ shown in Fig. \ref{fig:tgv_case2}.}
    \label{fig:enstrophy}
\end{figure}

Figure \ref{fig:tgv_case1} depicts the dissipation $dk/dt$ and the scaled enstrophy $\nu \mathcal{E}$ of the four focused velocity sets in comparison to the reference by Brachet et al. \cite{Brachet1983}. It is shown that all velocity sets match the reference solution for both dissipation and scaled enstrophy. Thus in case of well-resolved simulations, even the lean D3Q13 velocity set commended itself as a good choice for three-dimensional flow simulations.
In case of the underresolved simulations with Reynolds number $\mathrm{Re}=1600$ with time step size $\delta_t=0.00075$ the situation is slightly different though, displayed in Fig. \ref{fig:tgv_case2} showing different evolutions of dissipation and scaled enstrophy for each of the velocity sets. Initially, the dissipation followed the DNS reference simulation. As the flow entered the fully turbulent regime, the dissipation became more volatile than in the resolved case due to under-resolution of small vortices. In this phase, the numerical discretization provided dissipation in the spirit of an implicit LES \cite{Adams2009}. Despite the absence of an explicit subgrid model, all stencils produced macroscopic dissipation rates that roughly followed the reference solution. The D3Q13 and D3Q27 stencils even captured the location of the dissipation peak within 0.5 time units, whereas D3Q21 and D3V27 produced a plateau and a premature peak, respectively. That said, the broader implications of this difference are debatable, since the dissipation in the under-resolved regime is mostly a result of numerical errors interacting with the finest resolved scales.
In contrast, the scaled enstrophy is a more meaningful observable as it directly reflects the local shear stresses. 
 Fig. \ref{fig:enstrophy} compares the scaled enstrophies revealing that the enstrophy determined by simulations with the common D3Q27 most drastically underestimated the enstrophy in this configuration. Table \ref{tab:errors} confirms this observation by listing the Euclidean norm of the difference between the scaled enstrophies obtained by simulation and reference, i.e. 

\begin{equation}
    \|\nu \mathcal{E}^\mathrm{ref} -  \nu \mathcal{E}^\mathrm{sim}\| = \sqrt{\frac{1}{\hat{N}}\sum_{j=1}^{\hat{N}}\left(\nu \mathcal{E}^\mathrm{ref}_j -  \nu \mathcal{E}^\mathrm{sim}_j \right)^2},
\end{equation}
where $\hat{N}$ is the number of discrete measurements. The degree-seven D3V27 simulations as well as the degree-five D3Q21 simulations both more accurately predicted the scaled enstrophy. 


\begin{table}
    \centering
    \caption{Root mean squared error of the scaled enstrophy with respect to the reference \cite{Brachet1983}.}
    \label{tab:errors}

\begin{tabular}{ l c c c c} 
& D3Q13 & D3Q21 & D3V27 & D3Q27 \\ 
\hline
 $\|\nu \mathcal{E}^\mathrm{ref} -  \nu \mathcal{E}^\mathrm{sim}\|$  & 0.00141 & 0.00116 & 0.00107 & 0.00158 \\
\end{tabular}
\end{table}

\subsection{Compressible 3D Taylor-Green vortex}\label{sec:TGV3Dfully}

As a final test case the compressible 3D Taylor-Green vortex was simulated. Recently, this test case has been by extensively investigated by Peng and Yang \cite{Peng2018}.  The Reynolds number was $Re=400$, the heat capacity ratio was $\gamma=1.4$, and the Prandtl number was $\mathrm{Pr}=0.75$. The initial velocity field corresponds to Eq. \eqref{eq:TGV3D-velocity}, whereas the pressure is obtained by $P=\rho T_0$ with $T_0=1$, while the density is given by
\begin{equation}
    \rho(\bm{x},t=0) = 1.0 + \frac{\mathcal{C}^2}{16} (\mathrm{cos}(2x) + \mathrm{cos}(2y)) \mathrm{cos}(2z+2).
\end{equation}
This relation of temperature and density corresponds to the constant temperature initial condition (CTIC) detailed in Peng and Yang \cite{Peng2018}. The nominator $\mathcal{C}$ can be either set to $\mathcal{C}=\mathrm{Ma}^2$ or to $\mathcal{C}=1$; this work made use of the latter case. Table \ref{tab:setup-full3Dvortex} lists the most important simulation variables and Table \ref{tab:timestepsizes} separately lists the time step sizes and relaxation times at $\theta = 1.0$ of the respective simulations with the velocity sets D3Q103, D3V107, and D3Q45. The time step sizes of the on-lattice velocity sets were dictated by the configuration, whereas the time step sizes of the D3Q45 SLLBM simulation were adjustable.  The results of the D3Q45 velocity set is discussed in more depth in a follow-up manuscript \cite{Wilde2020a}. 

\begin{table*}
    \centering
    \caption{Simulation setup of the fully compressible 3D Taylor-Green vortex.}
    \label{tab:setup-full3Dvortex}
\begin{tabular}{ l c c c c c} 
Test case & & a) & b) & c) & d) \\ 
\hline
  Reynolds number & $\mathrm{Re}$ & 400 & 400 & 400 & 400 \\
  Mach number & $\mathrm{Ma}$ & 0.5 & 1.0 & 1.5 & 2.0  \\
  Spatial resolution & $\delta_x$  & 0.025 & 0.025 & 0.025 & 0.025 \\ 
  Truncation order of equilibrium & $N$ & 4 & 4 & 4 & 4 \\
  \hline
\end{tabular}
\end{table*}

\begin{figure}
    \centering
    \includegraphics{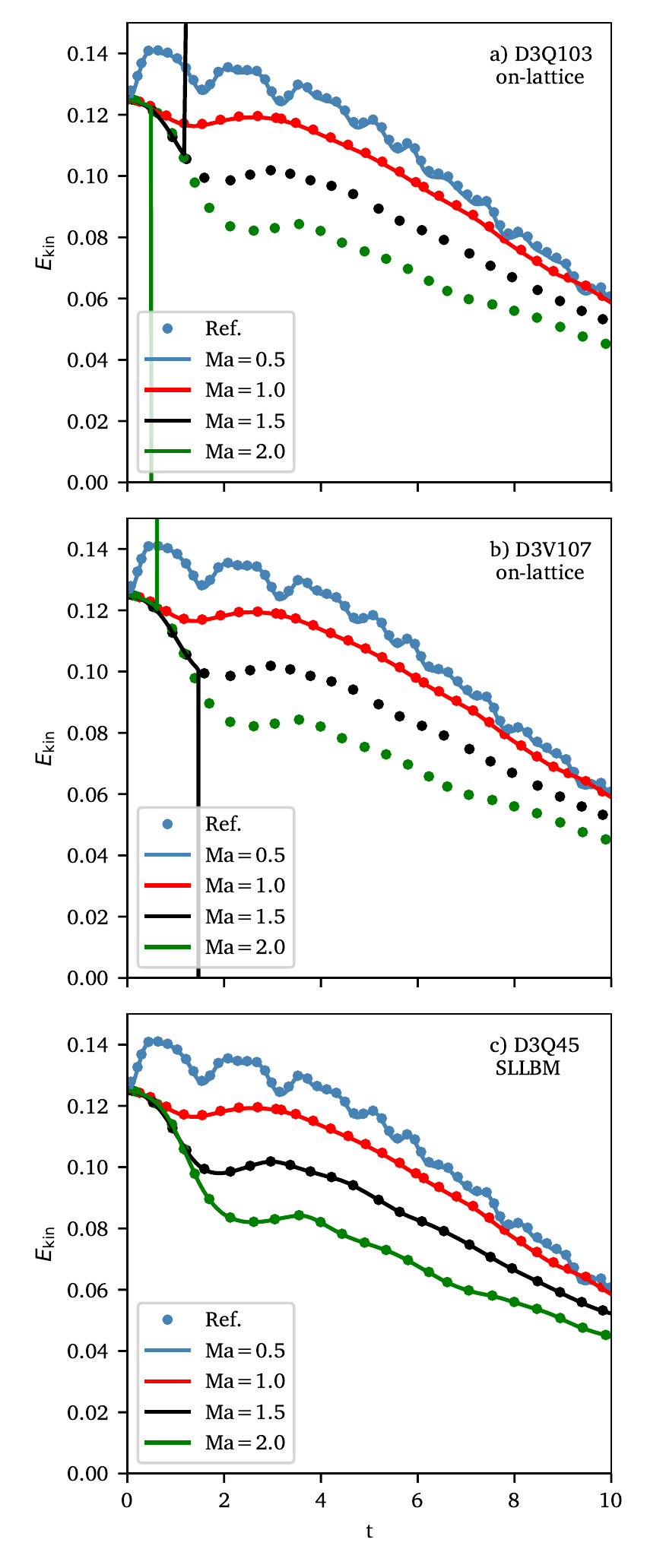}
    \caption{On-lattice simulations utilizing the D3Q103 a) and the D3V107 in comparison to the SLLBM D3Q45 simulation c). All velocity sets accurately reproduced the kinetic energies for the medium Mach numbers $\mathrm{Ma}=0.5$ and $\mathrm{Ma}=1.0$, while the on-lattice simulations failed for the higher Mach numbers $\mathrm{Ma}=1.5$ and $\mathrm{Ma}=2.0$. By contrast, the flexible time step size enabled successful SLLBM simulations even at the highest Mach numbers. }
    \label{fig:Compressible_TGV}
\end{figure}

Figure \ref{fig:Compressible_TGV} depicts the mean kinetic energy over time for a resolution of $256^3$ obtained by the velocity sets D3Q103, D3V107 and by the SLLBM D3Q45 simulation. It displays that the reference by Peng and Yang was well captured for Mach numbers $\mathrm{Ma}=0.5$ and $\mathrm{Ma}=1.0$ by all three velocity sets. However, the on-lattice simulations became unstable for the larger Mach numbers $\mathrm{Ma}=1.5$ and $\mathrm{Ma}=2.0$. 

These results confirm the assumptions of other works which predicted the capability of the on-lattice velocity sets D3Q103 and D3V107 to simulate compressible flows \cite{Shan2016,Surmas2009,Shan2010,Coreixas2018}. However, simulations applying these velocity sets lack stability, at least in the present configuration for Mach numbers exceeding one. In contrast, the variable time step size of the SLLBM enabled stable simulations even at higher Mach numbers using a substantially smaller velocity set with 45 discrete velocities.

\begin{table}
    \centering
    \caption{Time step sizes (top) and relaxation times at $\theta=1.0$ (bottom) of the compressible Taylor-Green vortex simulations. Unstable simulations marked by asterisks.}
    \label{tab:timestepsizes}
\begin{tabular}{ c c c c } 
  a) Time step size $\delta_t$
    & D3Q103 & D3V107 & D3Q45  \\
  \hline
  $\mathrm{Ma}=0.5$ & 0.0121 & 0.0136 & 0.0165  \\
  $\mathrm{Ma}=1.0$  & 0.0243 & 0.0271 & 0.0329 \\
  $\mathrm{Ma}=1.5$  & 0.0364* &  0.0407* & 0.0178 \\ 
  $\mathrm{Ma}=2.0$  & 0.0485* & 0.0542* & 0.0119
\end{tabular}

\begin{tabular}{ c c c c } 
  \\
  b) Relaxation time $\tau$ 
    & D3Q103 & D3V107 & D3Q45  \\
  \hline
  $\mathrm{Ma}=0.5$ & 0.572 & 0.564  & 0.553  \\
  $\mathrm{Ma}=1.0$  & 0.644 & 0.629  & 0.606 \\
  $\mathrm{Ma}=1.5$  & 0.716* & 0.694* &  0.943 \\ 
  $\mathrm{Ma}=2.0$  & 0.789* & 0.758* & 1.681
\end{tabular}
\end{table}

\section{Discussion} \label{sec:discussion}
The discretization of the velocity space by quadrature is a key asset of the lattice Boltzmann method. However, the coupling of the velocity space discretization with the spatial and temporal discretization in on-lattice Boltzmann methods is obstructive to find sufficiently small velocity sets especially for high quadrature order. Even a recent systematic study as done by Spiller and Dünweg did not yield smaller velocity sets than the already known D3Q103 \cite{Spiller2020}, whose number of abscissae therefore appears to be the lower limit at this quadrature degree with equidistant nodes. The enormous size of these high-degree sets prevented their application in actual simulations, at least up to the present study. The simulations of the fully compressible Taylor-Green vortex in the present work showed on-lattice velocity sets are generally capable for compressible simulations, but they lack stability at Mach numbers beyond unity. The biggest issue, from our point of view, is the fixed time step size of most compressible on-lattice Boltzmann methods. An exception to this are hybrid lattice Boltzmann methods, which solve the energy equation by finite volume or finite difference methods \cite{Feng2016,Feng2019}. These methods are capable to adjust the time step size by changing the reference temperature, but they suffer from restrictions in the stability regions \cite{Renard2020ALS}. Despite the successes of recent works \cite{Frapolli2015,Saadat2019, Frapolli2016a,Atif2018,Latt2020}, the applicability of compressible on-lattice Boltzmann methods remains vague.  Although shifted stencils have proven to extend the Mach number range of on-lattice Boltzmann methods \cite{Frapolli2016, Hosseini2019} in certain situations, the Taylor-Green vortex at high Mach numbers would still not be stable with static reference frames. Solely dynamic shifts of the reference frame as presented by Coreixas and Latt \cite{Coreixas2020a} might be an alternative for on-lattice Boltzmann methods, although they cause additional computational costs and require further investigation.  

Compared to on-lattice Boltzmann methods, the discretization of the velocity space is a largely unexplored field of research in off-lattice Boltzmann methods. We were able, however, to identify a large potential in equipping off-lattice Boltzmann methods by velocity sets with significantly different shapes. Instead of deriving stencils case-by-case, the research on multivariate quadrature rules yielded suitable cubature rules as templates for velocity sets.  Consequently, this work's main purpose was to explore these velocity sets by off-lattice Boltzmann simulations. 

The results in Section \ref{sec:results} clearly indicate that the spatial
 freedom obtained in the collocation of abscissae, as done in cubature rules, reduces the number of discrete velocities significantly. Simultaneously, the computational costs halved, narrowing the gap between on-lattice and off-lattice Boltzmann methods in efficiency. For example, the D2Q19 degree-nine velocity set is approximately half the size of its on-lattice counterpart D2V37 with equal quadrature order \cite{Philippi2006}. The difference between the D3Q45 and the D3Q103 velocity set is even larger, but the size of the D3Q45 is also superior compared to a recently utilized degree-nine D3Q77 off-lattice velocity set for compressible finite volume LBMs \cite{Guo2020}. The D3Q45 simulations of the Taylor-Green vortex proved to be stable and accurate even for high Mach number flows. This can be attributed to the temporal discretization of off-lattice Boltzmann methods, which is independent of the discretization of space and velocity space. 

Differences were also identified in terms of accuracy. The D3Q13 and D3Q21 based on icodahedra and dodecahedra, shown in Fig.~\ref{fig:platonics}, proved to be both efficient and accurate off-lattice velocity sets for weakly compressible flows. Both velocity sets showed a better enstrophy agreement with the reference solution in underresolved flow simulations. The reason for this is not necessarily that the precision of the cubature is higher, since the degree of precision of the D3Q13, D3Q21, and D3Q27 velocity sets is identical. Instead, we attribute the better agreement of the enstrophy with the reference to the fewer advection equations that have to be solved connected with fewer interpolation steps, leading to less artificial diffusion. Moreover, due to their derivation from platonic solids, these velocity sets exhibit a high geometric isotropy, which is favourable for the simulation of turbulent flows. This is probably also the reason why the D3Q21 outperforms the D3Q13 velocity set in terms of enstrophy. However, simulations with all of these degree-five velocity sets are affected by the cubic error term, being dependent on the Mach number. Obviously, this error term also spoils the simulations, since the Mach number of Ma=0.1 is not negligible. The degree-seven velocity set D3V27 with the aim to eliminate these cubic errors in the model proved to be more accurate than the on-lattice D3Q27, but coming at a comparable price. 
Despite the numerical diffusion introduced by the SLLBM, we showed in past works that the spatial convergence order corresponds to the order of the finite element \cite{Kramer2020}. By contrast, the spatial convergence order of on-lattice Boltzmann methods is restricted to second order.

Numerical diffusion and dispersion are critical issues of off-lattice Boltzmann methods compared to on-lattice Boltzmann methods, resulting from the interpolation in case of the SLLBM. High-order interpolation polynomials as used in the present article, however, significantly lower the dissipative effects in the simulation \cite{Kramer2020}. Contrary to that, on-lattice Boltzmann methods are valued for low dissipation \cite{Marie2009} due to the exact streaming step, but they suffer from instabilities at high Reynolds numbers when applying the usual BGK collision operator. To encounter these stability issues, a number of collision models have been proposed to stabilize simulations by introducing minimal diffusion without sacrificing the locality of the collision operator. Examples are multi-relaxation time (MRT)  \cite{Qian1992,DHumieres2002,Dellar2003, Shan2007}, central MRT \cite{Geier2006,Shan2019}, entropic \cite{Karlin1999,Atif2017}, regularized \cite{Latt2006,Malaspinas2015}, or cumulant models \cite{Geier2015,Geier2020}. Due to the non-negligible amount of dissipation evoked by the streaming step of off-lattice Boltzmann schemes, it is likely that these established on-lattice collision models fulfil a different role in off-lattice LBM. Nevertheless, previous work has utilized the semi-Lagrangian streaming step in combination with a stabilized collision model \cite{DiIlio2018}. Consequently, future research is necessary to clarify the impact of collision models on off-lattice Boltzmann methods.

Finally, some technical aspects need to be discussed. In particular, the streaming step’s implementations of on-lattice and off-lattice Boltzmann methods are significantly different. On the one hand, the exact streaming step of on-lattice Boltzmann methods can naively be implemented by on-board routines of many software packages, e.g. by “numpy.roll” in Numpy \cite{Numpy}. On the other hand, the interpolation routines of the SLLBM require a thoughtful implementation, especially when following our recommendation to employ high-order polynomials with non-equidistant support points and cell-wise organization. For this task, a matured finite element library like deal.ii is helpful \cite{dealII92}. The execution time of the semi-Lagrangian streaming step is one magnitude larger \cite{Kramer2020} for the same configuration and velocity set.. Despite the computational overhead, off-lattice Boltzmann methods have proven beneficial for unstructured meshes or compressible simulations \cite{DiIlio2018,Kramer2020,Wilde2020}. More appropriate velocity sets as presented in this article further reduce the gap in computation cost, as the computational complexity of the SLLBM scales linearly with respect to the number of discrete velocities.

To sum up, off-lattice Boltzmann methods relax the LBM in terms of the velocity space discretization. As the development of sparse numerical cubatures is still progressing, the present work lays the foundations to directly utilize new cubature rules in CFD simulations with off-lattice Boltzmann methods. 

\section{Conclusion} \label{sec:conclusion}
This paper studied the utilization of cubature rules in off-lattice Boltzmann methods for weakly and fully compressible flows. The deduced off-lattice velocity sets presented in this article reduce the number of discrete velocities and increase the accuracy of the simulation. In addition, fully compressible off-lattice simulations with degree-nine velocity sets feature better stability compared to the corresponding on-lattice counterparts.
Taken together, cubature rule-derived off-lattice velocity sets are a viable alternative to the customary on-lattice Boltzmann velocity sets and should have priority in off-lattice Boltzmann simulations.

\section*{Acknowledgements}
We gratefully acknowledge support for D.W. by German Research Foundation (DFG) project FO 674/17-1. The simulations were performed using the Platform for Scientific Computing at Bonn-Rhein-Sieg University, which is funded by the German Ministry of Education and Research and the Ministry for Culture and Science North Rhine-Westfalia (research grant 13FH156IN6).

\appendix 
\section{Hermite polynomials and moments}

The scaled Hermite polynomials with $\hat{\xi_i} = \xi_i/c_s$ up to fourth order read 

\begin{align*}
    {\mathcal{H}_i}^{(0)}&= 1 \\
    {\mathcal{H}_{i\alpha}}^{(1)}&=
                   \frac{{{\hat{\xi}_{i\alpha}}}}{c_s}  \\
   {\mathcal{H}_{i\alpha \beta}}^{(2)}&= \frac{{{{\hat{\xi}_{i\alpha}}{\hat{\xi}_{i\beta}}}}-\delta_{\alpha\beta}}{c_s^2} \\
    {\mathcal{H}_{i\alpha \beta \gamma}}^{(3)}&= \frac{{\hat{\xi}_{i\alpha}}{\hat{\xi}_{i\beta}}{\hat{\xi}_{i\gamma}}-({\hat{\xi}_{i\alpha}\delta_{\beta\gamma}}+{\hat{\xi}_{i\beta}\delta_{\alpha\gamma}}+{\hat{\xi}_{i\gamma}\delta_{\alpha\beta}})}{c_s^3} \\ 
    {\mathcal{H}_{i\alpha \beta \gamma \delta}}^{(4)}&= \frac{{{\hat{\xi}_{i\alpha}}{\hat{\xi}_{i\beta}}{\hat{\xi}_{i\gamma}}{\hat{\xi}_{i\delta}}}-\mathcal{T}_i+(\delta_{\alpha\beta}\delta_{\gamma\delta}+\delta_{\alpha\gamma}\delta_{\beta\delta}+\delta_{\alpha\delta}\delta_{\beta\gamma})}{c_s^4},
\end{align*}

with

\begin{equation*}
\begin{split}
            \mathcal{T}_i = \hat{\xi}_{i\alpha}\hat{\xi}_{i\beta}\delta_{\gamma\delta} + \hat{\xi}_{i\alpha}\hat{\xi}_{i\gamma}\delta_{\beta\delta} + \hat{\xi}_{i\alpha}\hat{\xi}_{i\delta}\delta_{\beta\gamma} + \\ \hat{\xi}_{i\beta}\hat{\xi}_{i\gamma}\delta_{\alpha\delta} + \hat{\xi}_{i\beta}\hat{\xi}_{i\delta}\delta_{\alpha\gamma}+ \hat{\xi}_{i\gamma}\hat{\xi}_{i\delta}\delta_{\alpha\beta}.
            \end{split}
\end{equation*}

The moments of the Boltzmann equation up to fourth order are

\begin{subequations}
\label{moments}
\begin{eqnarray*}
    a^{(0)}_{eq} &=& \rho \\
    a^{(1)}_{\alpha,eq} &=& \rho u_\alpha \\ 
    a^{(2)}_{\alpha \beta,eq} &=&  \rho (u_\alpha u_\beta + T_0(\theta -1 )\delta_{\alpha\beta}) \\
    a^{(3)}_{\alpha \beta \gamma,eq} &=& \rho \left[u_\alpha u_\beta u_\gamma +  T_0(\theta -1 )(\delta_{\alpha\beta}u_\gamma  \right.\nonumber\\
     & & \left.\hspace{1.2cm} + \delta_{\alpha\gamma}u_\beta +\delta_{\beta\gamma}u_\alpha)\right] \label{moments_Q}\\
    a^{(4)}_{\alpha \beta \gamma \delta,eq} &=&  \rho[ u_\alpha u_\beta u_\gamma u_\delta   
     + T_0(\theta-1)((\delta_{\alpha\beta}\delta_{\gamma\delta} \nonumber\\
     & &+ \delta_{\alpha\gamma}\delta_{\beta\delta}+ \delta_{\alpha\delta}\delta_{\beta\gamma} )T_0(\theta-1)  \nonumber\\ 
   & & + \delta_{\alpha\beta} u_{\gamma}u_{\delta} + \delta_{\alpha\gamma} u_{\beta}u_{\delta} + \delta_{\alpha\delta} u_{\beta}u_{\gamma} +\nonumber\\
   & & \delta_{\beta\gamma} u_{\alpha}u_{\delta} + \delta_{\beta\delta} u_{\alpha}u_{\gamma} + \delta_{\gamma\delta} u_{\alpha}u_{\beta}  )]. 
\vspace{.3cm}
\end{eqnarray*}
\end{subequations}

\section{Symmetry conditions}\label{app:symmetry} 

The symmetry conditions from zeroth to ninth order are

\begin{align*}
    \sum_i w_i &=1  \\
    \sum_i w_i \xi_{i\alpha} &=0 \\ 
    \sum_i w_i \xi_{i\alpha} \xi_{i\beta} &= c_s^2 \delta_{\alpha \beta} \\
    \sum_i w_i \xi_{i\alpha} \xi_{i\beta} \xi_{i\gamma} &=0 \\
    \sum_i w_i \xi_{i\alpha} \xi_{i\beta} \xi_{i\gamma} \xi_{i\delta}&= c_s^4 \Delta_{\alpha \beta \gamma \delta}  \\
    \sum_i w_i \xi_{i\alpha} \xi_{i\beta} \xi_{i\gamma} \xi_{i\delta} \xi_{i\epsilon}&= 0 \\
    \sum_i w_i \xi_{i\alpha} \xi_{i\beta} \xi_{i\gamma} \xi_{i\delta} \xi_{i\epsilon} \xi_{i\zeta}&= c_s^6\Delta_{\alpha \beta \gamma \delta \epsilon \zeta} \\
        \sum_i w_i \xi_{i\alpha} \xi_{i\beta} \xi_{i\gamma} \xi_{i\delta} \xi_{i\epsilon} \xi_{i\zeta}  \xi_{i\eta} &= 0 \\
    \sum_i w_i \xi_{i\alpha} \xi_{i\beta} \xi_{i\gamma} \xi_{i\delta} \xi_{i\epsilon} \xi_{i\zeta}  \xi_{i\eta} \xi_{i\theta} &= c_s^8\Delta_{\alpha \beta \gamma \delta \epsilon \zeta \eta \theta} \\
    \sum_i w_i \xi_{i\alpha} \xi_{i\beta} \xi_{i\gamma} \xi_{i\delta} \xi_{i\epsilon} \xi_{i\zeta}  \xi_{i\eta} \xi_{i\theta} \xi_{i\iota} &= 0,
\end{align*}
with
\begin{equation*}
        \Delta_{\alpha \beta \gamma \delta}=\delta_{\alpha \beta} \delta_{\gamma \delta} +\delta_{\alpha \gamma} \delta_{\beta \delta} +\delta_{\alpha \delta} \delta_{\beta \gamma},
\end{equation*}
as well as
\begin{multline*}
        \Delta_{\alpha \beta \gamma \delta \epsilon \zeta }=\delta_{\alpha \beta} \Delta_{\gamma \delta \epsilon \zeta} \\+\delta_{\alpha \gamma} \Delta_{\beta \delta \epsilon \zeta} +\delta_{\alpha \delta}\Delta_{\beta \gamma \epsilon \zeta}  +\delta_{\alpha \epsilon}\Delta_{\beta \gamma \delta \zeta} +\delta_{\alpha \zeta}\Delta_{\beta \gamma \delta \epsilon},
\end{multline*}
and
\begin{multline*}
        \Delta_{\alpha \beta \gamma \delta \epsilon \zeta \eta \theta}=\delta_{\alpha \beta} \Delta_{\gamma \delta \epsilon \zeta \eta \theta} +\delta_{\alpha \gamma} \Delta_{\beta \delta \epsilon \zeta \eta \theta}  +\delta_{\alpha \delta}\Delta_{\beta \gamma \epsilon \zeta \eta \theta}  +\delta_{\alpha \epsilon}\Delta_{\beta \gamma \delta \zeta \eta \theta} \\ +\delta_{\alpha \zeta}\Delta_{\beta \gamma \delta \epsilon \eta \theta} +\delta_{\alpha \eta}\Delta_{\beta \gamma \delta \epsilon \zeta \theta} +\delta_{\alpha \theta}\Delta_{\beta \gamma \delta \epsilon \zeta \eta}.
\end{multline*}

%BEGIN SECTION
\begin{table*}[h]
\section{D2Q19 velocity set} \label{app:d2q19}

\begin{tabular}{cccc}

$i$ & $w$ & $\xi_x$ & $\xi_y$ \\\hline
$0$ & 0.3168437267921905 & 0.0 & 0.0 \\
$1,2$ & 0.10558878375062891 & $\pm 1.4869982213169028 $ & 0.0 \\
$3,4,5,6 $ & 0.1024247123210936 & $\pm0.775196278121181$ & $\pm1.367469636752619$ \\

$7,8,9,10$ & 0.00953510698543825 & $\pm2.5175897644357486$ & $\pm1.105629214668943$ \\
$11,12$ & 0.006865104210104631 & $0.0$ & $\pm2.9213306655318734$ \\

$13,14,15,16$ &0.002405335328939458 & $\pm 1.8663975507141328$ & $\pm2.6987507639352253$ \\

$17,18$ & 0.0003939393722285871 & $\pm 3.8358342053914734$ & $0.0$ \\ \hline
$c_s = 1$
\end{tabular}

\end{table*}

\begin{table*}[h]
\section{D3V27 velocity set} \label{app:d3v27}

\begin{tabular}{cccccc}
$i$ & $w$ & $\xi_x$ & $\xi_y$ & $\xi_z$ \\\hline
$0$ & 0.31247897198654906 & 0.0 & 0.0 & 0.0 \\
$1,\ldots,8$ & 0.06338446047675325 & $\pm 1.1198362860638005$ & $\pm 1.1198362860638005$ &  $\pm$ 1.1198362860638005 \\
$9 ,\ldots,14$&0.029035130153906134 & $\pm 2.358709038202103$ & 0.0 & 0.0 & (cyc)   \\
$15 ,\ldots, 26 $& 0.0005195469396656799 & $\pm 3.142130383387586$ & $\pm 3.142130383387586 $&  0.0 & (cyc)  \\

$c_s = 1$
\end{tabular} 
\end{table*}

\begin{table*}
\section{D3Q45 velocity set} \label{app:d3q45}
\begin{tabular}{cccc}
$w$ & $\xi_x$ & $\xi_y$ & $\xi_z$ \\\hline
0.20740740740740618 & 0.0 & 0.0 & 0.0 \\
0.05787037037037047 & 0.06386083877343968 & -1.2239121278243665 & -1.2239121278243665 \\
0.05787037037037047 & -0.06386083877343968 & 1.2239121278243665 & 1.2239121278243665 \\
0.05787037037037047 & 1.2239121278243665 & -0.06386083877343968 & 1.2239121278243665 \\
0.05787037037037047 & -1.2239121278243665 & 0.06386083877343968 & -1.2239121278243665 \\
0.05787037037037047 & 1.2239121278243665 & 1.2239121278243665 & -0.06386083877343968 \\
0.05787037037037047 & -1.2239121278243665 & -1.2239121278243665 & 0.06386083877343968 \\
0.05787037037037047 & 1.5766994272507744 & -0.5069610024977665 & -0.5069610024977665 \\
0.05787037037037047 & -1.5766994272507744 & 0.5069610024977665 & 0.5069610024977665 \\
0.05787037037037047 & 0.5069610024977665 & 0.5069610024977665 & -1.5766994272507744 \\
0.05787037037037047 & -0.5069610024977665 & -0.5069610024977665 & 1.5766994272507744 \\
0.05787037037037047 & -0.5069610024977665 & 1.5766994272507744 & -0.5069610024977665 \\
0.05787037037037047 & 0.5069610024977665 & -1.5766994272507744 & 0.5069610024977665 \\
0.00462962962962958 & 2.403092127540177 & 0.8892242114059369 & -1.5602655313772367 \\
0.00462962962962958 & -2.403092127540177 & -0.8892242114059369 & 1.5602655313772367 \\
0.00462962962962958 & -2.403092127540177 & 1.5602655313772367 & -0.8892242114059369 \\
0.00462962962962958 & 2.403092127540177 & -1.5602655313772367 & 0.8892242114059369 \\
0.00462962962962958 & -0.8892242114059369 & -2.403092127540177 & 1.5602655313772367 \\
0.00462962962962958 & 0.8892242114059369 & 2.403092127540177 & -1.5602655313772367 \\
0.00462962962962958 & -0.8892242114059369 & 1.5602655313772367 & -2.403092127540177 \\
0.00462962962962958 & 0.8892242114059369 & -1.5602655313772367 & 2.403092127540177 \\
0.00462962962962958 & -1.5602655313772367 & 0.8892242114059369 & 2.403092127540177 \\
0.00462962962962958 & -1.5602655313772367 & 2.403092127540177 & 0.8892242114059369 \\
0.00462962962962958 & 1.5602655313772367 & -0.8892242114059369 & -2.403092127540177 \\
0.00462962962962958 & 1.5602655313772367 & -2.403092127540177 & -0.8892242114059369 \\
0.00462962962962958 & 0.4744978678080795 & 0.4744978678080795 & 2.9239876105912574 \\
0.00462962962962958 & 0.4744978678080795 & 2.9239876105912574 & 0.4744978678080795 \\
0.00462962962962958 & -0.4744978678080795 & -0.4744978678080795 & -2.9239876105912574 \\
0.00462962962962958 & -0.4744978678080795 & -2.9239876105912574 & -0.4744978678080795 \\
0.00462962962962958 & 2.9239876105912574 & 0.4744978678080795 & 0.4744978678080795 \\
0.00462962962962958 & -2.9239876105912574 & -0.4744978678080795 & -0.4744978678080795 \\
0.00462962962962958 & 1.7320508075688787 & 1.7320508075688787 & 1.7320508075688787 \\
0.00462962962962958 & -1.7320508075688787 & -1.7320508075688787 & -1.7320508075688787 \\
0.0004629629629629939 & -2.7367507163016924 & 0.14279717659756475 & -2.7367507163016924 \\
0.0004629629629629939 & 2.7367507163016924 & 2.7367507163016924 & -0.14279717659756475 \\
0.0004629629629629939 & 2.7367507163016924 & -0.14279717659756475 & 2.7367507163016924 \\
0.0004629629629629939 & -2.7367507163016924 & -2.7367507163016924 & 0.14279717659756475 \\
0.0004629629629629939 & 0.14279717659756475 & -2.7367507163016924 & -2.7367507163016924 \\
0.0004629629629629939 & -0.14279717659756475 & 2.7367507163016924 & 2.7367507163016924 \\
0.0004629629629629939 & -3.5256070994177073 & 1.1335992635264445 & 1.1335992635264445 \\
0.0004629629629629939 & 3.5256070994177073 & -1.1335992635264445 & -1.1335992635264445 \\
0.0004629629629629939 & 1.1335992635264445 & -3.5256070994177073 & 1.1335992635264445 \\
0.0004629629629629939 & -1.1335992635264445 & 3.5256070994177073 & -1.1335992635264445 \\
0.0004629629629629939 & 1.1335992635264445 & 1.1335992635264445 & -3.5256070994177073 \\
0.0004629629629629939 & -1.1335992635264445 & -1.1335992635264445 & 3.5256070994177073 \\ \hline
$c_s = 1$
\end{tabular} 
\end{table*}
%% \section{}
%% \label{}

%% If you have bibdatabase file and want bibtex to generate the
%% bibitems, please use
%%
%%  \bibliographystyle{elsarticle-num} 
%%  \bibliography{<your bibdatabase>}

%% else use the following coding to input the bibitems directly in the
%% TeX file.

\bibliography{jcs}
\bibliographystyle{unsrt}
\end{document}